\documentstyle[psfig]{l-aa}
\newhelp\stablestylehelp{You must choose a style between 0 and 3.}%
\newhelp\stablelinehelp{You should not use special hrules when stretching
a table.}%
\newhelp\stablesmultiplehelp{You have tried to place an S-Table inside another
S-Table.  I would recommend not going on.}%
\newdimen\stablesthinline
\newdimen\stablesthickline
\newif\ifstablesborderthin
\stablesborderthinfalse
\newif\ifstablesinternalthin
\stablesinternalthintrue
\newif\ifstablesomit
\newif\ifstablemode
\newif\ifstablesright
\stablesrightfalse
\newdimen\stablesbaselineskip
\newdimen\stableslineskip
\newdimen\stableslineskiplimit
\newcount\stablesmode
\newcount\stableslines
\newcount\stablestemp
\newcount\stablescount
\newcount\stableslinet
\newcount\stablestyle
\def\stablesleft{\quad\hfil}%
\def\stablesright{\hfil\quad}%
\catcode`\|=\active%
\newcount\stablestrutsize
\newbox\stablestrutbox
\setbox\stablestrutbox=\hbox{\vrule height10pt depth5pt width0pt}
\def\stablestrut{\relax\ifmmode%
                         \copy\stablestrutbox%
                       \else%
                         \unhcopy\stablestrutbox%
                       \fi}%
\newdimen\stablesborderwidth
\newdimen\stablesinternalwidth
\newdimen\stablesdummy
\newcount\stablesdummyc
\newif\ifstablesin
\stablesinfalse
\def\begintable{\stablestart%
  \stablemodetrue%
  \stablesadj%
  \halign%
  \stablesdef}%
\def\stablesadj{%
  \ifcase\stablestyle%
    \hbox to \hsize\bgroup\hss\vbox\bgroup%
  \or%
    \hbox to \hsize\bgroup\vbox\bgroup%
  \or%
    \hbox to \hsize\bgroup\hss\vbox\bgroup%
  \or%
    \hbox\bgroup\vbox\bgroup%
  \else%
    \errhelp=\stablestylehelp%
    \errmessage{Invalid style selected, using default}%
    \hbox to \hsize\bgroup\hss\vbox\bgroup%
  \fi}%
\def\stablesend{\egroup%
  \ifcase\stablestyle%
    \hss\egroup%
  \or%
    \hss\egroup%
  \or%
    \egroup%
  \or%
    \egroup%
  \else%
    \hss\egroup%
  \fi}%
\def\stablestart{%
  \ifstablesin%
    \errhelp=\stablesmultiplehelp%
    \errmessage{An S-Table cannot be placed within an S-Table!}%
  \fi
  \global\stablesintrue%
  \global\advance\stablescount by 1%
  \message{<S-Tables Generating Table \number\stablescount}%
  \begingroup%
  \stablestrutsize=\ht\stablestrutbox%
  \advance\stablestrutsize by \dp\stablestrutbox%
  \ifstablesborderthin%
    \stablesborderwidth=\stablesthinline%
  \else%
    \stablesborderwidth=\stablesthickline%
  \fi%
  \ifstablesinternalthin%
    \stablesinternalwidth=\stablesthinline%
  \else%
    \stablesinternalwidth=\stablesthickline%
  \fi%
  \tabskip=0pt%
  \stablesbaselineskip=\baselineskip%
  \stableslineskip=\lineskip%
  \stableslineskiplimit=\lineskiplimit%
  \offinterlineskip%
  \def\borderrule{\vrule width \stablesborderwidth}%
  \def\internalrule{\vrule width \stablesinternalwidth}%
  \def\thinline{\noalign{\hrule height \stablesthinline}}%
  \def\thickline{\noalign{\hrule height \stablesthickline}}%
  \def\trule{\omit\leaders\hrule height \stablesthinline\hfill}%
  \def\ttrule{\omit\leaders\hrule height \stablesthickline\hfill}%
  \def\tttrule##1{\omit\leaders\hrule height ##1\hfill}%
  \def\stablesel{&\omit\global\stablesmode=0%
    \global\advance\stableslines by 1\borderrule\hfil\cr}%
  \def\el{\stablesel&}%
  \def\elt{\stablesel\thinline&}%
  \def\eltt{\stablesel\thickline&}%
  \def\elttt##1{\stablesel\noalign{\hrule height ##1}&}%
  \def\elspec{&\omit\hfil\borderrule\cr\omit\borderrule&%
              \ifstablemode%
              \else%
                \errhelp=\stablelinehelp%
                \errmessage{Special ruling will not display properly}%
              \fi}%
  \def\stmultispan##1{\mscount=##1 \loop\ifnum\mscount>3 \stspan\repeat}%
  \def\stspan{\span\omit \advance\mscount by -1}%
  \def\multicolumn##1{\omit\multiply\stablestemp by ##1%
     \stmultispan{\stablestemp}%
     \advance\stablesmode by ##1%
     \advance\stablesmode by -1%
     \stablestemp=3}%
  \def\multirow##1{\stablesdummyc=##1\parindent=0pt\setbox0\hbox\bgroup%
    \aftergroup\emultirow\let\temp=}
  \def\emultirow{\setbox1\vbox to\stablesdummyc\stablestrutsize%
    {\hsize\wd0\vfil\box0\vfil}%
    \ht1=\ht\stablestrutbox%
    \dp1=\dp\stablestrutbox%
    \box1}%
  \def\stpar##1{\vtop\bgroup\hsize ##1%
     \baselineskip=\stablesbaselineskip%
     \lineskip=\stableslineskip%
     \lineskiplimit=\stableslineskiplimit\bgroup\aftergroup\estpar\let\temp=}%
  \def\estpar{\vskip 6pt\egroup}%
  \def\stparrow##1##2{\stablesdummy=##2%
     \setbox0=\vtop to ##1\stablestrutsize\bgroup%
     \hsize\stablesdummy%
     \baselineskip=\stablesbaselineskip%
     \lineskip=\stableslineskip%
     \lineskiplimit=\stableslineskiplimit%
     \bgroup\vfil\aftergroup\estparrow%
     \let\temp=}%
  \def\estparrow{\vfil\egroup%
     \ht0=\ht\stablestrutbox%
     \dp0=\dp\stablestrutbox%
     \wd0=\stablesdummy%
     \box0}%
  \def|{\global\advance\stablesmode by 1&&&}%
  \def\|{\global\advance\stablesmode by 1&\omit\vrule width 0pt%
         \hfil&&}%
  \def\vt{\global\advance\stablesmode by 1&\omit\vrule width \stablesthinline%
          \hfil&&}%
  \def\vtt{\global\advance\stablesmode by 1&\omit\vrule width \stablesthickline%
          \hfil&&}%
  \def\vttt##1{\global\advance\stablesmode by 1&\omit\vrule width ##1%
          \hfil&&}%
  \def\vtr{\global\advance\stablesmode by 1&\omit\hfil\vrule width%
           \stablesthinline&&}%
  \def\vttr{\global\advance\stablesmode by 1&\omit\hfil\vrule width%
            \stablesthickline&&}%
  \def\vtttr##1{\global\advance\stablesmode by 1&\omit\hfil\vrule width ##1&&}%
  \stableslines=0%
  \stablesomitfalse}
\def\stablesdef{\bgroup\stablestrut\borderrule##\tabskip=0pt plus 1fil%
  &\stablesleft##\stablesright%
  &##\ifstablesright\hfill\fi\internalrule\ifstablesright\else\hfill\fi%
  \tabskip 0pt&&##\hfil\tabskip=0pt plus 1fil%
  &\stablesleft##\stablesright%
  &##\ifstablesright\hfill\fi\internalrule\ifstablesright\else\hfill\fi%
  \tabskip=0pt\cr%
  \ifstablesborderthin%
    \thinline%
  \else%
    \thickline%
  \fi&%
}%
\def\endtable{\advance\stableslines by 1\advance\stablesmode by 1%
   \message{- Rows: \number\stableslines, Columns:  \number\stablesmode>}%
   \stablesel%
   \ifstablesborderthin%
     \thinline%
   \else%
     \thickline%
   \fi%
   \egroup\stablesend%
\endgroup%
\global\stablesinfalse}
\newcommand{\be}{\begin{equation}}
\newcommand{\en}{\end{equation}}
\begin{document}
% definitions ----------------------------------------------------
\def\ltsima{$\; \buildrel < \over \sim \;$}
\def\lsim{\lower.5ex\hbox{\ltsima}}
\def\gtsima{$\; \buildrel > \over \sim \;$}
\def\gsim{\lower.5ex\hbox{\gtsima}}
\def\spose#1{\hbox to 0pt{#1\hss}}
\def\approxlt{\mathrel{\spose{\lower 3pt\hbox{$\sim$}}
        \raise 2.0pt\hbox{$<$}}}
\def\approxgt{\mathrel{\spose{\lower 3pt\hbox{$\sim$}}
        \raise 2.0pt\hbox{$>$}}}
\def\deg {^\circ}
\def\mdot {\dot M}
\def\kms {$\sim$km$\sim$s$^{-1}$}
\def\gs {$\sim$g$\sim$s$^{-1}$}
\def\ergs {$\sim$erg$\sim$s$^{-1}$}
\def\cmtre {$\sim$cm$^{-3}$}\def\nupa{\vfill\eject\noindent}
\def\der#1#2{{d #1 \over d #2}}
\def\l#1{\lambda_{#1}}
\def\grb{$\gamma$-ray burst}
\def\grbs{$\gamma$-ray bursts}
\def\rosat{{\sl ROSAT} }
\def\cmdue {cm$^{-2}$}
\def\gcm {$\sim$g$\sim$cm$^{-3}$}
\def\rsole{$\sim$R_{\odot}}
\def\msole{$\sim$M_{\odot}}
\def\aa #1 #2 {A\&A, {#1}, #2}
\def\mon #1 #2 {MNRAS, {#1}, #2}
\def\apj #1 #2 {ApJ, {#1}, #2}
\def\nat #1 #2 {Nature, {#1}, #2}
\def\pasj #1 #2 {PASJ, {#1}, #2}
\newfont{\mc}{cmcsc10 scaled\magstep2}
\newfont{\cmc}{cmcsc10 scaled\magstep1}
\newcommand{\bc}{\begin{center}}
\newcommand{\ec}{\end{center}}
%----------------------------------

\title{$BeppoSAX$ observations of a new X--ray burster in the 
Galactic Center region, possibly coincident with a recurrent transient.}
%\subtitle
\author{L.~Sidoli\inst{1,\,2}, S.~Mereghetti\inst{1}, G.L.~Israel\inst{3,\,4}, 
G.~Cusumano\inst{5}, L.~Chiappetti\inst{1}, A.~Treves\inst{6}}

\institute{
{Istituto di Fisica Cosmica del C.N.R., Via Bassini 15, I-20133 Milano,
Italy; \\ e-mail: (sidoli, sandro, lucio)@ifctr.mi.cnr.it}
\and
{Dipartimento di Fisica, Universit\`a di Milano, Via Celoria 16, I-20133
Milano, Italy}
\and
{Osservatorio Astronomico di Roma, Via dell'Osservatorio 2,
I-00040 Monteporzio Catone (Roma), 
Italy; \\ e-mail: israel@coma.mporzio.astro.it}
\and
{Affiliated to I.C.R.A.}
\and
{Istituto di Fisica Cosmica ed Applicazioni all'Informatica del C.N.R., 
Via La Malfa 153, I-90146 Palermo, 
Italy; \\ e-mail: cusumano@ifcai.pa.cnr.it}
\and
{Universit\`a di Milano, sede di Como, Via Lucini 3, I-22100
Como, Italy; \\ e-mail: treves@astmiu.uni.mi.astro.it}
}

\maketitle
\label{sampout}

\begin{abstract}

We report BeppoSAX NFI observations of the   X--ray
source SAX~J1747.0--2853 recently discovered in the
region of the Galactic Center. 
The presence of type I X--ray bursts indicates that this
source, positionally coincident with the transient GX~0.2--0.2
observed in 1976, is a neutron star accreting from a 
low mass companion.

\keywords{Stars: neutron, individual: SAX~J1747.0--2853 -- X--rays: bursts}

\end{abstract}
 
\section{Introduction}

In the last few years repeated observations of the
galactic bulge region with coded mask hard X-ray 
telescopes have led to the discovery of many new sources.

The larger number of X-ray sources 
in the Galactic Center direction, compared to other parts 
of the galactic plane,
indicates the presence of an enhanced concentration of
accreting binaries in a region where the 
overall mass density is higher than at larger galactocentric 
distances.  Though most of these sources are basically of the
same kind of the accreting low mass and high mass  binaries 
found elsewhere in the Galaxy, a few of 
them   turned out to be particularly interesting and
peculiar objects , like e.g. the ``bursting pulsar'' GRO~J1744--28 
(Lewin et al. 1996), the ``microquasar'' 1E~1740.9--2942 (Mirabel et al. 1992),
and the 2 msec pulsar SAX J1808.4--3658 (in't Zand et al. 1998a, 
Wijnands \& van der Klis  1998). Here we report on the BeppoSAX 
MECS observation of a recently discovered 
bursting   X-ray source located at an angular distance of
19 arcmin from the Galactic Center direction. This source, 
named SAX~J1747.0--2853 , has been discovered  with the
Wide Field Camera instruments (WFC) on board BeppoSAX (in't Zand et al. 1998b)
and has been later observed with both 
the Narrow Field Instruments (NFI) and the WFC instruments by Bazzano
and collaborators (Bazzano et al. 1998). These authors also
reported the presence of X--ray bursts from SAX~J1747.0--2853. The data 
described here were obtained about 20 days after the
discovery of SAX~J1747.0--2853, as part of our survey of
the Galactic Center region with the BeppoSAX NFI (Sidoli et al. 1998).

\section{Observations and Data Analysis}

The region of sky containing SAX~J1747.0--2853 was imaged with 
the MECS and  LECS instruments during  an observation 
performed from April 13 to April 15, 1998.
The MECS instrument (Boella et al. 1997) is based on position-sensitive
gas-scintillation proportional counters providing images in the   
1.3-10 keV energy range within a field of view of 56 arcmin diameter.
After standard data selection and cleaning, the resulting 
net exposure time in the   MECS instrument is  72 ksec.
 
As 1E~1743.1-2843 was the main target of the observation, 
the new source  SAX~J1747.0--2853 was observed $\sim$ 13 arcmin
off axis, at coordinates 
$R.A.=17h~47m~0.5s, Dec.=-28\deg~52'~36''$, J2000 (with an uncertainty 
of $\sim$ 1 arcmin). This position is consistent
with that obtained   in the  discovery observation with the WFC instrument
($R.A.=17h~47m~02s, Dec.=-28\deg~52'.0$, J2000, 
$3'$ error radius, in't Zand et al. 1998)
and subsequently refined with the NFI observations 
($R.A.=17h~47m~02s, Dec.=-28\deg~52'.5$, J2000,
 $1'$ error radius) by Bazzano et al. (1998). in't Zand and coworkers 
 noted that the source 
SAX~J1747.0--2853 is positionally coincident, 
within errors, with the X--ray transient GX~0.2--0.2 observed in outburst
in 1976 (Proctor et al., 1978).

\begin{figure*}[!th]
%\vskip -6truecm
\vskip -7.5truecm
%\centerline{\psfig{figure=fig1.ps,width=10cm} {\hfil}}
\hskip 0.5truecm
\psfig{figure=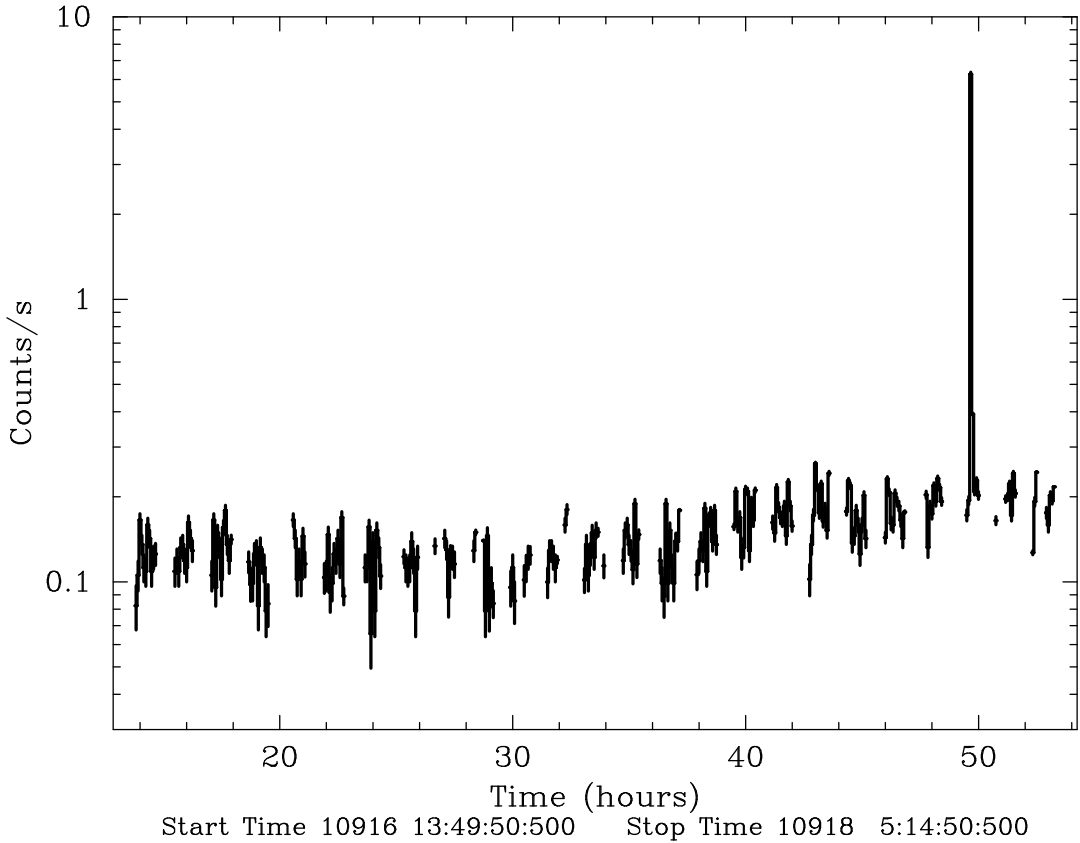,width=14cm} {\hfil}
\caption{SAX~J1747.0--2853 MECS light curve. The bin size is 200 s.}
\end{figure*}

Some  variability on a timescale of hours, as well as a strong burst, 
starting at 1:40:03 UT of 1998 April 15 (see section 3), 
are clearly visible in the MECS background
subtracted light curves presented in figure 1. After converting the 
arrival times to the Solar System barycenter, 
we searched for coherent signals in the X--ray flux of the source. 
We first removed the burst, then we accumulated a 0.3\,s binned light 
curve and calculated a single power spectrum over the whole 
observation following the method outlined by Israel \& Stella (1996). 
No significant periodicity was found with a 
corresponding 99\% confidence level pulsed fraction upper 
limits between 25\%--20\% and 15\%--17\% for the 10$^4$\,s--100\,s 
and 100\,s--1\,s period intervals, respectively. 

We also performed a 
period search during the 25s long burst time interval even in presence 
of poorer statistics. In this case we accumulated a 0.5\,ms 
light curve and calculated the corresponding power spectrum. We found 
no significant peak. The 99\% confidence level upper limits on the pulsed  
fraction are in the range 60\%--40\% and 35\%--40\% for the 1--500\,ms and 
500\,ms--2\,s period intervals, respectively.
  
To study the source spectrum we extracted the MECS counts 
within a radius of 2.5 arcmin from the source position
and rebinned them in order to have at
least 20 counts per bin. 
The time interval corresponding to
the X-ray burst was excluded from the analysis. 
To properly fit the count spectrum, we derived the MECS response matrix 
appropriate to
the source off-axis position and corrected for the adopted extraction radius.
The latter was smaller than the usual 4 arcmin in order to avoid 
regions of the detector affected by the support strongback of the
MECS window. 

The standard MECS background spectrum obtained from blank field observations
underestimates the actual background present in regions of low
galactic latitude. We therefore estimated  the background spectrum from 
the  region surrounding SAX~J1747.0--2853 observed in our data. 
The best fit 
with an absorbed power law gives a photon index $\alpha \sim$ 2.4
and an absorbing column density of $\sim$10$^{23}$ cm$^{-2}$. Though the 
formal uncertainties on these parameters are rather small, the results are somewhat 
dependent on the particular choice of the background region. 
Therefore, the errors indicated in Table 1, where the results of the fits are summarized,
have  been  estimated taking the background uncertainty into account.
The observed 2-10 keV flux  is  $\sim2\times 10^{-11}$ erg cm$^{-2}$ s$^{-1}$. A slightly  
better fit of the MECS spectrum can be obtained with a $\sim6$ keV thermal bremsstrahlung.

\begin{table*}
\label{spe}
\stablesthinline=0pt
\stablesborderthintrue
\stablestyle=0
\caption{{\bf} Persistent emission spectral analysis (errors are 90\% c.l.).}
\begintable
Model        | Column density        | parameter                           |Red. $\chi^2$   |Unabsorbed Flux (2--10 keV)         \el
             |($10^{22}$ cm$^{-2}$)  |                                     |(138 d.o.f.)    |($10^{-11}$ ergs cm$^{-2}$ s$^{-1}$)   \el
 
Power law    	|$9.9^{+0.8}_{-0.8}$   |$\alpha_{ph}=2.4^{+0.1}_{-0.1}$     | 1.09  |  $4.8^{+0.2}_{-0.3}$   \el
Bremsstrahlung  |$8.3^{+0.6}_{-0.3}$   |$T=6.1^{+0.9}_{-0.7}$ keV           | 1.01  |  $4.0^{+0.2}_{-0.3}$   \endtable
\end{table*}

\section{Analysis of the X-ray burst}

The light curve of the X-ray burst is shown in figure 2; 
only the orbit containing the burst is displayed. At the burst peak 
the source reaches a count rate $\sim500$ times stronger
than the persistent emission. By fitting the
burst light curves in two energy bands, we derived exponential
decay constants of $\sim$15 s and $\sim$10 s at energies respectively
below and above 4.5 keV. The average  spectrum of the burst 
is harder than that of the 
persistent emission (power law photon index $\sim$ 1.7). 
In figure 3 we show the results obtained by fitting the burst emission
in different time intervals with a blackbody spectrum.
The contribution from the persistent emission has been subtracted.
The spectral softening seen in the light curves is confirmed
by the spectral analysis that shows a temperature variation from 
$\sim$2 to $\sim$0.5 keV. Assuming an Eddington luminosity at the burst
peak, we obtain a distance of $\sim$ 10 kpc. 
Besides, assuming a spherical emitter at a distance of  $\sim$10 kpc,
 we determine a blackbody radius consistent with the canonical 
radius for a neutron star. Moreover, this value remains constant 
during the burst. 

\begin{figure}[!th]
%\vskip 9truecm
%%\centerline{\psfig{figure=figburst.ps,width=7cm} {\hfil}}
\centerline{\psfig{figure=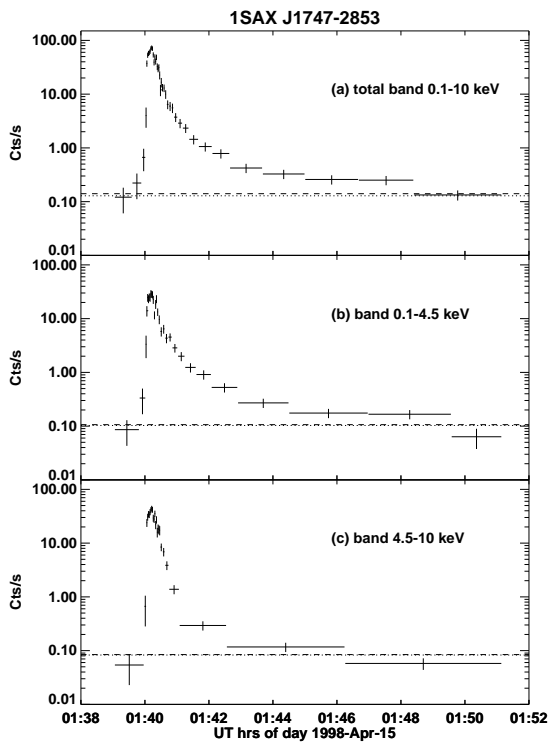,width=7cm} {\hfil}}
\caption{Light curve of the burst in the total, soft and hard bands (panels a,b,c
respectively). The minimum bin size is 1.5 s, but data have been rebinned for
clarity in such a way that each bin has at least a statistics of 5 sigma
during the burst decay (all bands), and a statistics of 2 $\sigma$ (panels a,b)
or 1.5 $\sigma$ (panel c) during the burst onset. The dotted and dashed lines 
indicate the level of the persistent emission in the contiguous interval 
(uninterrupted by Earth occultation or
SAGA passage) respectively before and after the burst.}
\end{figure}

\section{Discussion}

The properties of the   burst observed from SAX~J1747.0--2853 are typical
of type I X-ray bursts and allow to clearly classify this source as a 
neutron star in a Low Mass X-ray Binary. 
It is in fact widely believed that the type I 
X-ray bursts result from thermonuclear flashes on the 
surface of accreting neutron stars  (see, e.g., 
Maraschi \& Cavaliere 1977, Lewin et al. 1992).
 
\begin{figure}[!hbt]
%\vskip 9truecm
%%\centerline{\psfig{figure=figparam.ps,width=7cm} {\hfil}}
\centerline{\psfig{figure=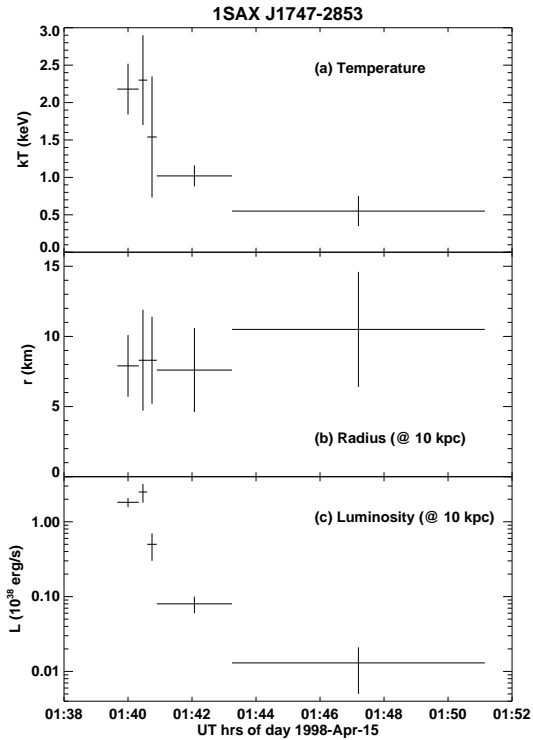,width=7cm} {\hfil}}
\caption{Evolution of temperature, luminosity and blackbody radius during
the SAX~J1747.0--2853 burst. For luminosity and radius a distance of 10 kpc 
was assumed.}
\end{figure}
  
Considering its galactic coordinates, the high absorption, 
and the properties of the burst, 
it is likely that    SAX~J1747.0--2853 be at a distance comparable to
that of the galactic center. 
For a distance of $\sim$ 10 kpc, also
the flux of   $\sim5\times 10^{-10}$ erg cm$^{-2}$ s$^{-1}$ observed  
during the previous BeppoSAX observations (Bazzano et al. 1998) is   
consistent with the typical luminosity of X-ray bursters 
(10$^{36}$ --10$^{37}$ erg s$^{-1}$).   

At the time of the observation we present here, the flux had decreased by about 
a factor ten with respect to the first NFI observation by Bazzano et al.(1998), 
consistent with  an exponential decay 
with  e-folding time of $\sim8$ days (Fig. 4). 
Unfortunately, the poor coverage of
the light curve does not allow to determine the shape of the outburst.

No other X-ray sources at this position have been reported previously,
with the exception of the transient   GX~0.2--0.2, that was active for a
few months in 1976 (Proctor, Skinner \& Willmore 1978, Cruddace et al. 1978).
GX~0.2--0.2 was observed with rocket-borne
instruments with limited angular resolution and more precisely located
with the RMC instrument on the Ariel V  satellite. The Ariel V position
of GX~0.2--0.2   (90\% confidence radius $\sim$ 1.5 arcmin)
is consistent with that of SAX~J1747.0--2853. 
Although it cannot be  excluded that different sources were   observed,
it is very likely that SAX~J1747.0--2853 and GX~0.2--0.2 are the same 
object, as already pointed out by in't Zand et al. (1998b). 
No bursts were observed from GX~0.2--0.2 in 1976, but the peak luminosity, the
duration of the ouburst and the relatively soft spectral shape 
(Proctor, Skinner \& Willmore 1978) were similar to those
observed in SAX~J1747.0--2853.

\begin{figure}[!hbt]
\vskip -2truecm
%%\centerline{\psfig{figure=curve_2.ps,width=7cm} {\hfil}}
%\centerline{\psfig{figure=f4.ps,width=6.5cm} {\hfil}}
\centerline{\psfig{figure=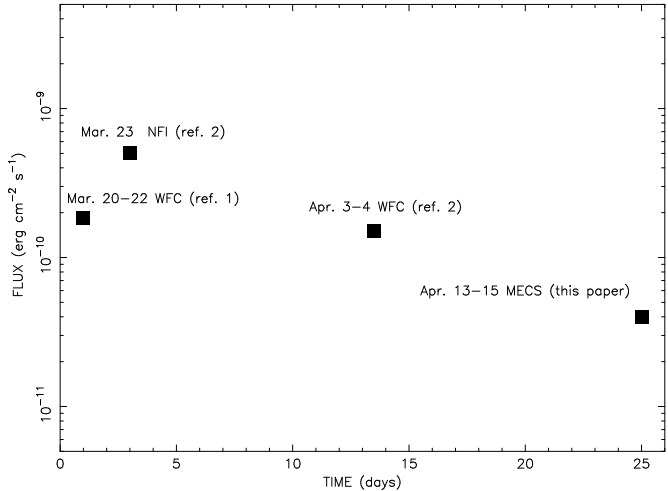,width=6.5cm} {\hfil}}
\caption{SAX~J1747.0--2853 recent BeppoSAX detections. 
The first three fluxes are taken from in't Zand (1998b, ref. 1) 
and Bazzano et al. (1998, ref 2.). 
In ref.1 and ref. 2 flux uncertainties are not reported; our errors are
smaller than the symbol used.}
\end{figure}

In figure 5 the fluxes observed in 1976 and 1998 are compared 
with several upper limits that we have derived from published
observations of the Galactic Center region.
All the fluxes have been converted to the 2-10 keV band assuming
the spectral parameters of our best fit. 
The  SIGMA observations (Goldwurm et al. 1994)
obtained at E$>$40 keV have not been reported since the corresponding
upper limits ($\sim0.6-3\times 10^{-9}$ erg cm$^{-2}$ s$^{-1}$) 
are not very constraining when extrapolated 
to lower energies with such a soft spectrum. 
For a different reason, i.e. the strong interstellar absorption, also
the ROSAT data are not very useful to constrain the luminosity 
of SAX~J1747.0--2853.

\begin{figure}[!htb]
\vskip -2.4truecm
%\centerline{\psfig{figure=f5new.ps,width=6.5cm} {\hfil}}
%\centerline{\psfig{figure=f5.ps,width=6.5cm} {\hfil}}
\centerline{\psfig{figure=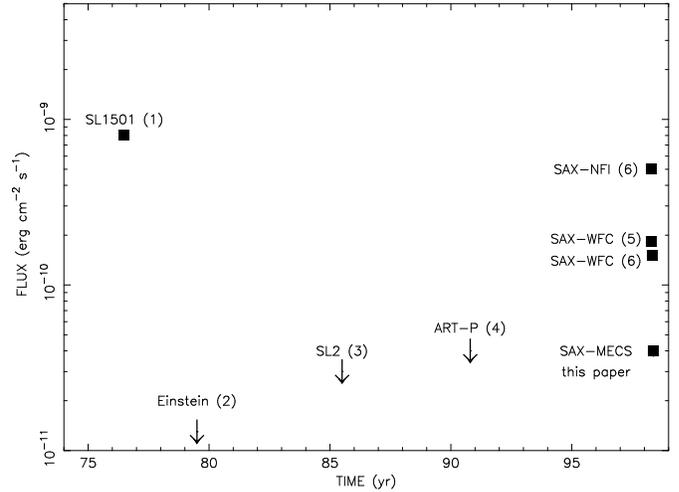,width=6.5cm} {\hfil}}
\caption{SAX~J1747.0--2853 detections and upper limits from 1975 to 1998.
References are: (1) Proctor et al. 78; (2) Watson et al. 81; 
(3) Skinner et al. 87; (4) Pavlinsky et al. 94; (5) in't Zand et al. 98b; 
(6) Bazzano et al. 98.}
\end{figure}

\section{Conclusions}

Using the MECS on--board BeppoSAX we detected a type I X--ray burst 
from the recently discovered source SAX~J1747.0--2853. 
The estimated position supports the identification with the transient
GX 0.2-0.2 observed in 1976. 
The X--ray burst is clearly indicative of the presence
of a neutron star accreting matter from a low mass companion.
The analysis of the burst properties allow to estimate a distance 
to the source of $\sim$ 10 kpc,
that place SAX~J1747.0--2853 close to the Galactic Center.
The severe interstellar absorption in this region hampers the search  
for the optical counterpart. 
Applying the relation 
$N(HI+H_{2})$/$A_{V}$  = $1.9\times 10^{21}$ cm$^{-2}$ mag$^{-1}$  
(Bohlin, Savage and Drake 1978), 
we estimate an extinction of  $\sim$50 mag, which means an apparent J magnitude
$>$30 for the low mass companion.

\begin{acknowledgements}
Lara Sidoli thanks Annamaria Borriello for help and useful discussions.
  
\end{acknowledgements}

\end{document}